\documentclass{article}
\usepackage{cite}
\usepackage{array}
\usepackage{amsmath,amsfonts,amssymb,bm,algorithm,mathabx}
\usepackage{graphicx}
\usepackage[colorlinks=true, allcolors=blue]{hyperref}
\usepackage[noend]{algorithmic}
\usepackage[caption=false,font=normalsize,labelfont=sf,textfont=sf]{subfig}

\usepackage{hyperref}
\hypersetup{
	unicode,
	pdfauthor={Author One, Author Two, Author Three},
	pdftitle={Quantum Annealing for Robust Principal Component Analysis},
	pdfsubject={Quantum Annealing for Robust Principal Component Analysis},
	pdfkeywords={Principal Component Analysis, Quantum Annealing, Dimensionality Reduction, Robust Subspace Learning},
	pdfproducer={LaTeX},
	pdfcreator={pdflatex}
}

\graphicspath{{./Figures/}}

\usepackage{textcomp}
\usepackage{stfloats}
\usepackage{url}
\usepackage{verbatim}

\def\BibTeX{{\rm B\kern-.05em{\sc i\kern-.025em b}\kern-.08em
    T\kern-.1667em\lower.7ex\hbox{E}\kern-.125emX}}

\DeclareMathOperator*{\argmax}{arg\,max}
\DeclareMathOperator*{\argmin}{arg\,min}

\begin{document}

\title{Quantum Annealing for Robust Principal Component Analysis}

\author{Ian Tomeo$^1$ \and Panos P. Markopoulos$^2$ \and Andreas Savakis$^1$}

\date{
	$^1$Rochester Institute of Technology,  Rochester NY, USA \\ \texttt{iet6395@rit.edu}\\%
	$^2$The University of Texas at San Antonio, San Antonio TX, USA \\ \texttt{panagiotis.markopoulos@utsa.edu}\\[2ex]%
	Date of publication Dec 15, 2024, date of current version Jan 25, 2025.
}

%\tfootnote{This research was partly supported by Air Force Office of Scientific Research (AFOSR) grant FA9550-20-1-0039.}

%\markboth

%\corresp{Corresponding author: Panos P. Markopoulos (panagtiotis.markopoulos@utsa.edu).}

\maketitle

\begin{abstract}{
Principal component analysis is commonly used 
for dimensionality reduction, feature extraction, denoising, and visualization. The most commonly used principal component analysis method is based upon optimization of the L2-norm, however, 
the L2-norm is known to exaggerate the contribution of errors and outliers. When optimizing over the L1-norm, the components generated are known to exhibit robustness or resistance to outliers in the data.  The L1-norm components can be solved for with a binary optimization problem. Previously, L1-BF has been used to solve the binary optimization for multiple components simultaneously. In this paper we propose QAPCA, a new method for finding principal components using quantum annealing hardware which will optimize over the robust L1-norm. The conditions required for convergence of the annealing problem are discussed. The potential speedup when using quantum annealing is demonstrated through complexity analysis and experimental results. To showcase performance against classical  principal component analysis techniques experiments upon synthetic Gaussian data, a fault detection scenario and breast cancer diagnostic data are studied. We find that the reconstruction error when using QAPCA is comparable to that when using L1-BF.}
\end{abstract}

\section{Introduction}
\label{sec:intro}
Principal component analysis (PCA) \cite{pearson1901liii} is a fundamental method for machine learning, data analysis, and parameter estimation. PCA is often used as a pre-processing technique when performing the tasks of data compression, fault detection, or classification. 
PCA is known to improve the utility of data by reducing noise and providing structure by only retaining the most informative features.
Efforts have been made to attribute known properties to components, imposing a structure upon the subspace. Kernel PCA (KPCA) \cite{scholkopf1998nonlinear} utilizes a kernel function to achieve nonlinear PCA. KPCA constructs a nonlinear map from the input space to the feature space and this allows components to represent nonlinear features. In sparse PCA (SPCA) \cite{zou2006sparse}, increasing sparsity can lead to better data interpretability because a sparse projection allows individual components to represent unique features of the data set.

\subsection{Robust PCA}
The robustness of PCA estimates is very important in feature creation. Because the standard PCA is sensitive to outliers,  techniques have been developed to confer robustness to the PCA estimates. For instance, L1-norm PCA method was developed and proposed \cite{markopoulos2017efficient, markopoulos2014optimal, GripPCA} replaces the squared emphasis that standard PCA allocates to each data point by a linear emphasis. Thus, the contribution of all data points is balanced and the solution more robust. In $R_1$-PCA \cite{ding2006r} a robust subspace is created that maintains rotational invariance. 

\subsection{Quantum Computing and Annealing}
Quantum computers hold the promise of fast and simple solutions for problems that are known to be highly complex when solved on classical computers. 
Modern integrated circuit fabrication has given rise to coherent optical and superconducting-based systems that have been configured to execute gate-model quantum computing. This type of computing has given rise to unique algorithms. Grover's Algorithm \cite{grover} can be used as an optimal database search, finding entries that satisfy a given constraint. Shor's Algorithm \cite{shor} can quickly factor prime numbers from an integer, and this has created a push for quantum resilient cryptography methods. A fast quantum computing based PCA is proposed in Lloyd\cite{lloyd2014quantum} by using density matrix exponentiation and this method was applied to a finance dataset for pricing of derivatives \cite{martin2021toward}.

An Adiabatic Quantum Computer has been optimized to accomplish what is known as Quantum Annealing (QA). These annealers are useful for solving combinatorial problems formatted to have binary solutions. There have been recent attempts to use these quantum annealers for classification, compression and feature selection. A support vector machine (SVM) formed on a QA was assigned to classify DNA sequences \cite{willsch2020support}. Classification of low-resolution images with deep learning is explored on a QA-based restricted Boltzmann machine (RBM) \cite{dixit2021training}. The compression of a statistical dataset with QA is demonstrated on a QCD lattice, a dataset of physics interactions\cite{yoon2022lossy}. Feature selection with QA is performed on the Indian Pine hyperspectral image (HSI) dataset, then QA based classification is executed \cite{otgonbaatar2021quantum}. Here, we propose the first strategy to carry out QA based PCA.

\subsection{Our Contributions}
The contributions of this work are related to the creation of Quantum Annealing Principle Component Analysis (QAPCA). A PCA method that runs on a quantum annealer; to the best of our knowledge, it is the first of its kind. The contributions are listed in the following bullets.

\begin{itemize}      
    \item The major contribution is we propose the first PCA method solved on quantum annealing hardware. Both a recursively solved single component solution and a faster method capable of computing multiple components simultaneously are proposed. The proposed QAPCA method, seeks to optimize the robust L1-PCA metric and, thus, inherits its robustness.
    \item We create an embedding process which relies upon a banding method to fit more samples on the hardware when performing an annealing operation. Intelligently reducing the number of couplers between qubits in this way can decrease the complexity of problems which assign equal importance between the qubits being annealed.
    \item We perform complexity analysis and comparison of QAPCA to similar classically solved algorithms. This highlights the various performance versus complexity tradeoff found in QAPCA.
    \item We test the efficacy of QAPCA on various numerical studies. A synthetic Gaussian dataset is used to test a hyperparameter of QAPCA. Detection accuracy of a corrupted fault detection scenario is explored. Also, the reconstruction error of a breast cancer detection dataset with mislabeled input data is tested.
\end{itemize}

\section{Methods}
\label{sec:background knowledge}

\subsection{PCA} 
PCA is extensively used for data analysis, including feature extraction and dimension reduction \cite{abdi2010principal}. With PCA we seek to reduce the original high dimension $D$ of data to $K<D$ truly informative features. Even in the case where there is only a small number of features, defining the most important ones helps to discern the salient characteristics of the data. From an algebraic standpoint, we are looking for a $K$-dimensional linear subspace that minimizes the representation error of the measurements. Thus, given dataset $\mathbf X \in R^{D\times N}$ containing $N$ samples, PCA is formulated using the L2-norm as follows:
\begin{align}
\hat{\mathbf R} = {
                \begin{smallmatrix}
                \argmax\\
\mathbf R \in \mathbb R^{D\times K},\mathbf R^\top\mathbf R=\mathbf I_K~ 
\end{smallmatrix}}
\|\mathbf R^\top \mathbf X\|_F^2,
\end{align} 
where $\| \cdot \|_F$ is the Frobenius (L2) norm of a matrix. We will let $\mathbf I_K$ represent an identity matrix of size $K$. The compressed dataset 
% that contains only the most informative linearly combined features 
is accordingly given by   $\mathbf Y = \hat{\mathbf R}^\top \mathbf X$. The PCA solution is commonly obtained by singular value decomposition (SVD) of $X$ \cite{golub1971singular}, or, equivalently, eigenvalue decomposition (EVD) of the estimated covariance matrix $\frac{1}{N}\mathbf X^\top \mathbf X$. The computational cost of SVD 
% in `` "classical" computing 
is $\mathcal O(\min\{N,D\}ND)$, and for huge datasets, SVD becomes impractical.

\subsection{L\textsubscript{1}-PCA}

When the data contains outliers (large-magnitude, 
out-of-distribution samples)
% peripheral, unexpected entries), 
standard L2-norm PCA does not perform well. The L1-norm PCA is a popular variant of PCA, that performs similarly to L2-PCA for clean data, while offering robustness to 
% also having the ability to suppress 
high-magnitude outliers.  
L1-PCA is formulated as 
\begin{align}
\hat{\mathbf R}_{L_1} = {
                \begin{smallmatrix}
                \argmax\\
\mathbf R \in \mathbb R^{D\times K},\mathbf R^\top \mathbf R= \mathbf I_K~ 
\end{smallmatrix}}
 \|\mathbf R^\top \mathbf X\|_1, 
\label{l1pca}
\end{align}
where $\| \cdot\|_1$ is the L1-norm of a vector (sum of absolute entries). The maximization of the sum of L1-norm components in \eqref{l1pca} promotes balanced emphasis of all data-points in the formulation of principal components, effectively suppressing outliers. 
Importantly, L1-PCA 
% has been proven admit a combinatorial reformulation 
has been formulated as a combinatorial optimization problem\cite{markopoulos2014optimal}.
It was shown that the solution to \eqref{l1pca} can be obtained by 
\begin{align}
\hat{\mathbf R}_{L_1}= \Phi (\mathbf X \mathbf B_{\text{opt}}),
\end{align}
where
\begin{align}
\mathbf B_\text{opt}= {
                \begin{smallmatrix}
                \argmax\\
\mathbf B \in \{ \pm 1 \}^{N \times K}
\end{smallmatrix}} \sum_{k=1}^K {\sigma_k [\mathbf B^\top \mathbf X^\top \mathbf X \mathbf B]}   \text{.}
\label{binarynuc}
\end{align} 
Here, $K$ is the number of PCA components, and $\sigma_k [\cdot]$ denotes the $k^{th}$ singular-value of its argument (in decreasing order). We express the binary matrix as $\mathbf B \in \{ \pm 1 \}^{N \times K}$. $\Phi (\cdot)$ returns the nearest orthonormal matrix to its argument, in the Euclidean sense. That is, for any tall matrix $\mathbf T$, we define
\begin{align}
\Phi(\mathbf T):={
                \begin{smallmatrix}
                \argmin\\
\mathbf G\in \mathbb R^{D\times K},\mathbf G^\top \mathbf G=\mathbf I_K
\end{smallmatrix}} \|\mathbf T - \mathbf G\|_F
\label{Procrustes}
\end{align} 
which is obtained by singular-value decomposition of $\mathbf T$ \cite{markopoulos2014optimal}.

\subsection{Quantum Information Processing}
Currently, quantum processing is emerging as a promising
approach for addressing computationally challenging problems. 
The seminal work by 
Lloyd\cite{lloyd2014quantum} proposed a quantum-processing method for PCA (qPCA). The theoretical speedup achieved can be enormous, as the requirements for performing SVD can be avoided through the use of specialized quantum hardware. Although, qPCA is theoretically very interesting, no hardware yet exists 
that could perform this technique with enough qubits to provide useful results in the field of big data. For example, recent results on the IBM Q quantum computer have shown that the utility of this technique is currently limited \cite{martin2021toward}. However, the potential speedups offered by quantum hardware are too useful to be ignored. 

\subsection{Quantum Annealing for Combinatorial Problems}
There exists hardware 
that leverages the process of quantum annealing, and efficient algorithms based on QA for solving combinatorial problems of very high complexity have been proposed\cite{venegas2018cross}. Such problems can be NP-hard, meaning there is no algorithm which can find a solution within polynomial time \cite{karp2010reducibility}. For example, QA can efficiently solve the class of Quadratic Unconstrained Binary Optimization (QUBO) \cite{QUBO} and Ising problems \cite{headquarters2020technical}, with the ability to utilize thousands of qubits.
For combinatorial problems that seek to optimize $M$ binary variables/weights, the number of possible combinations increases by $2^M$. Such combinatorial problems are typically NP-hard.
Algorithms to search for the proper binary weights typically make assumptions for reducing the search space, possibly causing significant decrease to the attainable performance
of the combinatorial problem. 

Importantly, QA can efficiently solve binary problems that are in Ising form
\begin{align}
\mathbf z_\text{opt}= {
                \begin{smallmatrix}
                \argmin\\
\mathbf z \in \{ -1,1\}^{N} \end{smallmatrix}}~ \mathbf z^T \mathbf J \mathbf z . 
\end{align}
An Ising spin glass model utilizes binary elements in its solution -1's and +1's defining the minimization. On a quantum annealer, the matrix $\mathbf J$ is represented by magnetic coupling strengths between qubits. Given a short time, the annealing system will settle with the set of qubits in a state that is the answer to the original minimization problem. By converting our L1-norm problem into a format that can fit on the QA hardware, we can load it on an annealer and quickly solve the NP-hard version of the problem without making any additional assumptions.

Using a quantum annealer for feature reduction has been presented before in QUBO format. The work in \cite{yoon2022lossy} is very similar to the QAPCA presented here. However, instead of a fast method for PCA, their work focuses on statistical data compression, so the samples are iteratively updated with the L-BFGS-B algorithm\cite{zhu1995l_bfgs_b} and subsequently rerun QA. The premise of selecting samples on the annealer has been used in \cite{mucke2023feature} where a similar solution controlled by an importance matrix masked samples to perform the feature reduction. QAPCA differs from the above methods in that we create robust components utilizing all of the available input samples when forming $\mathbf R$ while still suppressing the influence of outliers.
% such as fault detection. 
We have arrived at this solution from previous derivations of L1-PCA which naturally fit the Ising spin glass structure instead of using QUBO.

\section{Proposed Quantum Annealing PCA}
\subsection{Single Component L1-PCA in Ising Form}
 
We now show that, for a single component, L1-PCA can be straightforwardly brought into Ising form. Noticing that, for $K=1$, the expression in \eqref{binarynuc} becomes:
\begin{align}
\mathbf b_\text{opt}= {
                \begin{smallmatrix}
                \argmax\\
\hat{\mathbf b} \in \{ \pm 1 \}^{N}
\end{smallmatrix}} \|\mathbf X\hat{\mathbf b}\|_2 = {
                \begin{smallmatrix}
                \argmax\\
\hat{\mathbf b} \in \{ \pm 1 \}^{N}
\end{smallmatrix}}  \hat{\mathbf b}^\top \mathbf X^\top \mathbf X \hat{\mathbf b} \text{.}\label{ISING_FORMAT_A}
\end{align}

This binary quadratic form is formally an NP-hard problem. We can rewrite this formulation in Ising form as:
\begin{align}
\mathbf b_\text{opt}= {
                \begin{smallmatrix}
                \argmin\\
\hat{\mathbf b} \in \{ \pm 1 \}^{N}
\end{smallmatrix}} \hat{\mathbf b}^\top (-\mathbf X^\top \mathbf X) \hat{\mathbf b}= {
                \begin{smallmatrix}
                \argmin\\
\hat{\mathbf b} \in \{ \pm 1 \}^{N}
\end{smallmatrix}} \hat{\mathbf b}^\top \mathbf J \hat{\mathbf b} \text{.}
\label{ISING_FORMAT}
\end{align}

The model in \eqref{ISING_FORMAT} is already in Ising form with QA coupling strengths defined by $\mathbf J$. In \cite{tomeo2024l1} we solved this Ising problem by loading $\mathbf J$ onto the hardware the annealing process is run multiple times to find the $\hat{\mathbf b}$ which minimizes \eqref{ISING_FORMAT}.
 For multiple components $K>1$, we must perform a sequence of nullspace projections as $\mathbf X_k \leftarrow \mathbf X_{k-1} -  \mathbf r_{k-1} \mathbf r^\top_{k-1} \mathbf X_{k-1}$, $k > 1$, and repeat the annealing steps. This operation is recursively performed for $K$ components. We will denote this operation as QAPCA-R.
 
 %Here we acknowledge another method wherein QAPCA is only performed for a single component. Then, using this component we project its nullspace across $\mathbf J$. QAPCA is again performed upon the new subspace. 

\subsection{Multiple Components L1-PCA in Ising Form}
The aforementioned problem can benefit from a formulation that increases the number of simultaneously calculated components. By doing so, the PCA can be solved faster as the problem only needs to be annealed a single time instead of breaking the problem into multiple steps with nullspace projections. Additionally, a multiple-component QAPCA can be more accurate than a single component PCA, as the optimal solution to \eqref{binarynuc} does not preclude oblique components. 
We begin the reformulation by solving \eqref{binarynuc} directly.
\begin{subequations}\label{MultNucNorm1}
\begin{align}
\hat{\mathbf B} &= {
                \begin{smallmatrix}
                \argmax\\
                \mathbf B \in \{ \pm 1 \}^{N \times K}
                \end{smallmatrix}} \|\mathbf X \mathbf B \|_{*}^2   \label{MultNucNorm1a}\\
\hat{\mathbf B} &= {
                \begin{smallmatrix}
                \argmax\\
                \mathbf B \in \{ \pm 1 \}^{N \times K}
                \end{smallmatrix}} \text{Tr}(\sqrt{\mathbf B^\top \mathbf X^\top \mathbf X \mathbf B})^2 \text{.}\label{MultNucNorm1b}
\end{align}
\end{subequations}

%We will assume a bias $\mathbf H \in \mathbb R^{N \times K}$. 
\noindent
Using Cauchy-Schwartz, the first term in \eqref{MultNucNorm1a} can be simplified as 
\begin{subequations}
\label{MultNucNorm2}
\begin{align}
\text{Tr}(\sqrt{\mathbf B^\top \mathbf X^\top \mathbf X \mathbf B})^2 = 
\text{Tr}(\sqrt{\mathbf B^\top \mathbf X^\top \mathbf X \mathbf B} \mathbf I_K)^2 \label{MultNucNorm2a}\\
\text{Tr}(\sqrt{\mathbf B^\top \mathbf X^\top \mathbf X \mathbf B} \mathbf I_K)^2 \leq 
\text{Tr}(\sqrt{\mathbf B^\top \mathbf X^\top \mathbf X \mathbf B}^2)\text{Tr}(\mathbf I_K^2) \label{MultNucNorm2b}\\
\text{Tr}(\sqrt{\mathbf B^\top \mathbf X^\top \mathbf X \mathbf B}^2)\text{Tr}(\mathbf I_K^2) = K \text{Tr}(\mathbf B^\top \mathbf X^\top \mathbf X \mathbf B)
\text{.} \label{MultNucNorm2c}
\end{align}                

\end{subequations}
%\begin{align}
%    \text{Tr}(\sqrt{\mathbf B^\top \mathbf X^\top \mathbf X \mathbf B})^2 = \text{Tr}(\sqrt{\mathbf B^\top \mathbf X^\top \mathbf X \mathbf B} \mathbf I)^2 \leq \text{Tr}(\sqrt{\mathbf B^\top \mathbf X^\top \mathbf X \mathbf B}^2)\text{Tr}(\mathbf I^2) = K \text{Tr}(\mathbf B^\top \mathbf X^\top \mathbf X \mathbf B).
%    \label{MultNucNorm2}
%\end{align}
\noindent
The resultant term from \eqref{MultNucNorm2c} can again be simplified using the the binary component vectors $\mathbf b_k$, from the columns of $\mathbf B = [\mathbf b_1, \mathbf b_2, \dots, \mathbf b_K]$. Changing the problem into an argument minimization leads to
\begin{subequations}\label{MultNucNorm3}
\begin{align}
\hat{\mathbf B} &= {
                \begin{smallmatrix}
                \argmax\\
                \mathbf B \in \{ \pm 1 \}^{N \times K}
                \end{smallmatrix}} K\text{Tr}(\mathbf B^\top \mathbf X^\top \mathbf X \mathbf B)  \label{MultNucNorm3a}\\
\hat{\mathbf B} &= {
                \begin{smallmatrix}
                \argmin\\
                \mathbf B \in \{ \pm 1 \}^{N \times K}
                \end{smallmatrix}} K \sum_k \mathbf b_{k}^\top \mathbf J \mathbf b_{k} \text{.}\label{MultNucNorm3b}
\end{align}
\end{subequations}

In order to solve a binary problem on a quantum annealer we shall vectorize matrix $\mathbf B$ such that $\mathbf b' = \text{vec}(\mathbf B) = [\mathbf b_1^\top, \mathbf b_2^\top, \dots, \mathbf b_K^\top]^\top$. The problem in \eqref{MultNucNorm3} now takes the form of a large Ising problem as
\begin{align}
\hat{\mathbf b'} = {
                \begin{smallmatrix}
                \argmin\\
                \mathbf b' \in \{ \pm 1 \}^{KN}
                \end{smallmatrix}} \mathbf b'^\top K [\mathbf I_K \bigotimes \mathbf J] \mathbf b'.
                \label{MultNucNorm4}
\end{align}
 Here $\bigotimes$ is the Kronecker product \cite{kolda2009tensor} and it is being used to expand the problem to account for $K$ components. An issue with the formulation in \eqref{MultNucNorm4} is that there is no motivation for the problem to output different results for individual $\mathbf b_k$ other than adding an associated bias term. We propose that another reasonable way to ensure different results for different $\mathbf b_k$ is to introduce an orthogonality constraint (i.e., $\sum_{k_1 \neq k_2} \mathbf b_{k_1}^\top \mathbf J \mathbf b_{k_2} = 0$).

Adding a constraint which includes cross terms can be closer to the nuclear norm. If we let ${\mathbf 1}$ denote a vector of ones with size $K$ then $\epsilon {\mathbf 1}^T (\mathbf B^\top \mathbf X^\top \mathbf X \mathbf B) {\mathbf 1}$ will include cross terms.  Our goal is for our new constraint to be closer to the nuclear norm than the Frobenius norm. This will be shown as 
\begin{multline}
    \text{Tr}(\sqrt{\mathbf B^\top \mathbf X^\top \mathbf X \mathbf B})^2  \leq \\ (K + \epsilon) \text{Tr}(\mathbf B^\top \mathbf X^\top \mathbf X \mathbf B) - \epsilon {\mathbf 1}^T (\mathbf B^\top \mathbf X^\top \mathbf X \mathbf B) {\mathbf 1} \leq \\ K \text{Tr}(\mathbf B^\top \mathbf X^\top \mathbf X \mathbf B) \text{.}
    \label{EpsBound1}
\end{multline}

From \eqref{EpsBound1} we solve for $\epsilon$ by subtracting the leftmost term then dividing by $K \text{Tr}(\mathbf B^\top \mathbf X^\top \mathbf X \mathbf B) - \text{Tr}(\sqrt{\mathbf B^\top \mathbf X^\top \mathbf X \mathbf B})^2$. 
\begin{align}
     0 \leq 1 - \epsilon \frac{{\mathbf 1}^T (\mathbf B^\top \mathbf X^\top \mathbf X \mathbf B) {\mathbf 1}  -\text{Tr}(\mathbf B^\top \mathbf X^\top \mathbf X \mathbf B)}{K \text{Tr}(\mathbf B^\top \mathbf X^\top \mathbf X \mathbf B) - \text{Tr}(\sqrt{\mathbf B^\top \mathbf X^\top \mathbf X \mathbf B})^2} \leq  1.
    \label{EpsBound4}
\end{align}
Finally the equation is manipulated until we bound $\epsilon$ as
\begin{align}
      0 \leq \epsilon \leq \frac{K \text{Tr}(\mathbf B^\top \mathbf X^\top \mathbf X \mathbf B) - \text{Tr}(\sqrt{\mathbf B^\top \mathbf X^\top \mathbf X \mathbf B})^2}{{\mathbf 1}^T (\mathbf B^\top \mathbf X^\top \mathbf X \mathbf B) {\mathbf 1}  -\text{Tr}(\mathbf B^\top \mathbf X^\top \mathbf X \mathbf B)}.
    \label{EpsBound5}
\end{align} If $\mathbf X = \mathbf 1_{D \times N}$ and $\mathbf B = \mathbf 1_{N \times K}$ are all-ones matrices, we can solve \eqref{EpsBound5} and find our bound as 
\begin{align}
      \epsilon \leq \frac{K^2 N^2 D-K N^2 D}{K^2 N^2 D-K N^2 D} = 1.
    \label{EpsBound6}
\end{align}
%If $X \in \mathbb [-1,1]^{D \times N}$,
We can prove \eqref{EpsBound1} by looking at the terms in \eqref{EpsBound5}. Because the numerator in \eqref{EpsBound5} has been shown to be always positive in \eqref{MultNucNorm2} we find the denominator which allows the inequality
\begin{align}
    {\mathbf 1}^T (\mathbf B^\top \mathbf X^\top \mathbf X \mathbf B) {\mathbf 1} \geq \text{Tr}(\mathbf B^\top \mathbf X^\top \mathbf X \mathbf B)
    \label{EpsBound2}
\end{align}
 to hold. From \eqref{EpsBound2} we can now see
\begin{align}
     \sum_{k_1 \neq k_2} \mathbf b_{k_1}^\top \mathbf X^\top \mathbf X \mathbf b_{k_2} \geq 0.
    \label{EpsBound3}
\end{align}
Equation \eqref{EpsBound3} will typically hold if $\mathbf X^\top \mathbf X$ is positive semidefinite and cross terms are positively aligned. To show this, eigenvalue decomposition is performed on $\mathbf X^\top \mathbf X = \mathbf Q \mathbf \Lambda \mathbf Q^\top$ and with this we 
can form $\mathbf z_k = \mathbf Q^\top \mathbf b_k $. Here $\mathbf \Lambda = \text{diag}([\lambda_1,\dots,\lambda_n])$ is the diagonal matrix of eigenvalues and $\mathbf Q = [\mathbf q_1, \dots, \mathbf q_N]$ represents their relevant eigenvectors. We now use \eqref{EpsBound3} to form
\begin{align}
     \sum_{k_1 \neq k_2} \mathbf z_{k_1}^\top \mathbf \Lambda \mathbf z_{k_2} = \sum_{k_1 \neq k_2} \sum_{n} \lambda_{n} z_{n,k_1} z_{n,k_2} \geq 0.
    \label{EpsPD}
\end{align} If high energy components are aligned, equation \eqref{EpsPD} will hold. Because $\mathbf X^\top \mathbf X$ is positive semidefinite, the eigenvalues $\lambda_{n}$ will always be nonnegative and the sum of the components $z_{n,k}$ of $\mathbf z_k$ in the direction of the $n$-th eigenvector $\mathbf q_n$ will have $z_{n,k_1} z_{n,k_2} \geq 0$.

By introducing the orthogonality constraint in Equation \eqref{MultNucNorm3b} we have
\begin{align}
\hat{\mathbf B} = {
                \begin{smallmatrix}
                \argmin\\
                \mathbf b \in \{ \pm 1 \}^{N \times K}
                \end{smallmatrix} }(K+\epsilon) \sum_k \mathbf b_{k}^\top \mathbf J \mathbf b_{k} - \epsilon \sum_{k_1 \neq k_2} \mathbf b_{k_1}^\top \mathbf J \mathbf b_{k_2}\text{.}
                \label{MultNucNorm5}
\end{align}
This can then be put into Ising format as
\begin{align}
\hat{\mathbf b'} = {
                \begin{smallmatrix}
                \argmin\\
                \mathbf b' \in \{ \pm 1 \}^{KN}
                \end{smallmatrix}} \mathbf b'^\top [\mathbf I_K \bigotimes K \mathbf J + (\mathbf 1_K-\mathbf I_K) \bigotimes -\epsilon \mathbf J] \mathbf b'\text{.}
                \label{MultNucNorm6}
\end{align}
Here, $\mathbf 1_K \in \mathbf 1^{K \times K}$ is a square ones matrix of size $K$. If we use a strong enough orthogonality constraint (i.e., $\epsilon > 0$) we find that we do not need to impose a bias for the Ising problem to converge.

While L1-PCA is a good fit for QA realization, a quantum version of L2-PCA would suffer from various issues. One issue is that the L2 problem is not setup as a binary optimization problem, and setting it up in a binary form would cause the resulting PCA to have far less granularity than its L1 counterpart. Much like QAPCA, the QA L2-PCA would suffer from exponentially more qubits when $K>1$. When large enough quantum annealing hardware becomes available to mitigate these issues, a QA L2-PCA could perform faster than the SVD solution.

\subsection{Embedding}
D-Wave's \textit{Advantage} system \cite{mcgeoch2021advantage} can fit problems using $5,000$ qubits with $40,000$ couplers. For QAPCA, there must be a unique qubit corresponding to each of the $N$ samples. Each pair of qubits is 
coupled with a strength defined by an entry in $\mathbf J$. The usable amount of these qubits and couplers is determined by the chains required to fit a problem on the hardware. As each qubit on this hardware can attach to only 15 others, chains must be created to connect a qubit to more than 15 other entries in order to reach physically distant areas on an annealer. 
A chain can be formed by setting a coupler to full strength between one or more qubits, as this forms a single, large qubit from the system of qubits and couplers. Chained qubits/couplers 
represent a single unique entry in $\mathbf J$ for the Ising solution.

To maximize the number of entries of the $\mathbf J$ matrix that are represented on the quantum hardware  we need to utilize, as many of the $N^2$ couplers between different logical qubits as possible. 
Because $\mathbf J$ is symmetric, only entries from its upper triangular portion are needed to capture the unique aspects of a problem. Thus, only upper triangular entries of $\mathbf J$ are loaded onto the annealer. Entries which are shown to have zero correlation do not provide meaningful influence upon a problem's solution and can be discarded, or left unrepresented on the annealing hardware. If $K>1$, we also need to load $\mathbf J$ multiple times into the embedding. This means the number of unique couplers needed for a problem is \begin{align}
C = \frac{K^2N^2 - KN}{2} + KN\text{.}
\label{Couplers}
\end{align}
\noindent
Here, $C$ is the number of entries in $\mathbf J$ used to solve the problem. If all $C$ entries of $\mathbf J$ can be represented in an embedding, this is known as fully embedding the problem. 

In our studies on the \textit{Advantage} hardware, only solutions utilizing $N \leq N_\text{limit}$ could be fully embedded on the annealer. Plugging in 
% our max 
the maximum
number of couplers on the \textit{Advantage} system, i.e. 40,000, 
% (40,000) 
into Equation \eqref{Couplers} 
% we see that 
we can define a maximum number of training samples $N_\text{limit}$ which we can fully embed to be less than $300$ when $K = 1$. In practice we have found that on the \textit{Advantage} system we must have $N_\text{limit} \leq 175$ due to couplers lost when creating qubit chains \cite{mcgeoch2021advantage}. This allows us to define the maximum amount of usable couplers $C_\text{limit}$ to be defined by plugging $N_\text{limit}$ into Equation \eqref{Couplers}. Here, we must accounting for longer chains due to more samples by using a modified limit $N_{\text{limit}} - 25  = 150$ when calculating $C_{\text{limit}} = 11,325$.

\begin{figure}[t!]
	\centering
	\includegraphics[width=5in]{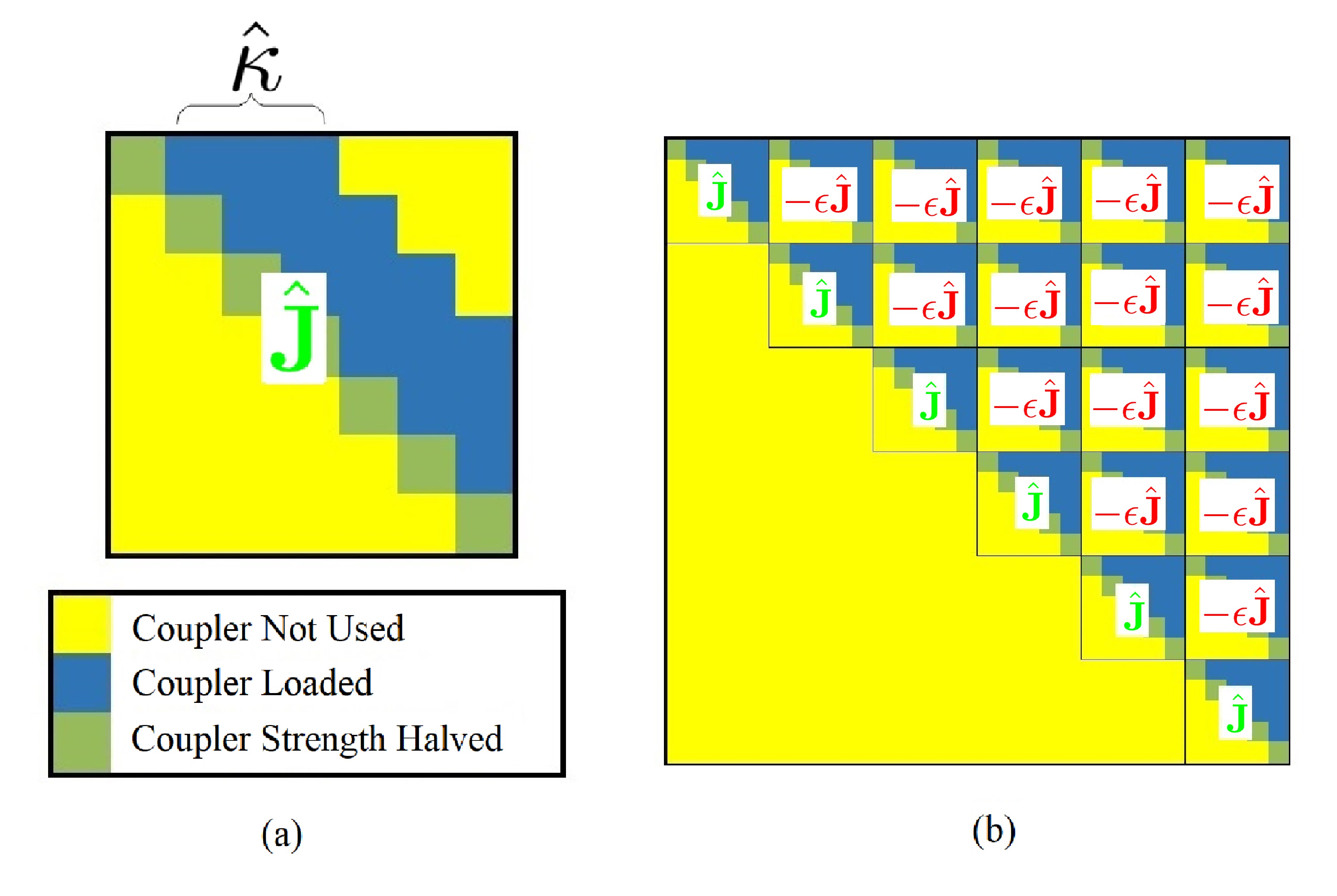}
	\caption{Example of the $\hat{\mathbf J}$ matrix used for QAPCA when (a) $N > N_\text{limit}$ (b) $K > 1$}
	\label{BandingandMultipleEmbedding}
\end{figure}

In our experiments, as an alternative to boosting \cite{freund1999short}, when $N > N_\text{limit}$ we begin instituting a banding procedure upon the matrix \cite{bickel2008regularized} as we can only achieve a partial embedding of the problem. To reduce the number of couplers required, we employ a banding technique to reduce qubit interconnectivity and thus reduce the amount of couplers used by deleting bands from the upper triangular matrix. To ensure proper convergence, every entry on the main diagonal which corresponds to a specific qubit must have at least 1 other nonzero entry in its corresponding row/column ensuring every sample is coupled to at least one other sample. We begin by loading the main diagonal of $\mathbf J$ into an estimate $\hat{\mathbf J}$ and then load subsequent diagonal bands from the upper triangular with an increasing offset $\kappa$ ensuring the number of couplers used is $ \leq C_\text{limit}$.
Once it is detected that adding another diagonal band will cause the number of couplers to exceed $C_\text{limit}$ the loading process is terminated and we map the resulting $\hat{\mathbf J}$ onto the annealer. When $K=1$, for row $i\text{ and column }j \in \{1, 2, \dots, N\}$,  we define
\begin{align}
    [\hat{\mathbf J}]_{i,j} :=  \left\{ \begin{matrix} [{\mathbf J}]_{i,j}, & \text{if } 0<j-i<\hat{\kappa}  \\ 1/2 ~[{\mathbf J}]_{i,j}, & \text{if } i=j  \\ 0, & \text{otherwise} \end{matrix}\right\} \text{,}
    \label{SingleJ}
\end{align}
where the offset is calculated as
\begin{align}    
\hat{\kappa} = \argmax_{
\begin{smallmatrix}
\kappa \in \{0, 1, \dots, N-1\}
\end{smallmatrix}}  \frac{1}{2} (2 N -\kappa)(\kappa+1) \leq C_\text{limit}.
\end{align} This embedding for a single component is shown in Figure \ref{BandingandMultipleEmbedding}(a) depicting how the banding procedure can affect the upper triangular matrix $\hat{\mathbf J}$. This technique will now be extended to embed multiple components. Equation \eqref{SingleJ} is used $\frac{K^2 - K}{2}$ times to create the embedding formation visualized in \ref{BandingandMultipleEmbedding}b. To form the multiple components embedding of $\mathbf J$ we use
\begin{multline}
    [\hat{\mathbf J}_{K>1}]_{(k_1 + 1) i + k_1 N, (k_2 + 1) j + k_2 N } :=  \\ \left\{ \begin{matrix} [{\mathbf J}]_{i,j}, & \text{if } 0<j-i<\hat{\kappa} \text{ and } k_1=k_2 \\ 1/2 ~[{\mathbf J}]_{i,j}, & \text{if } i=j \text{ and } k_1=k_2 \\ -\epsilon [{\mathbf J}]_{i,j}, & \text{if } 0<j-i<\hat{\kappa} \text{ and } k_1 \neq k_2 \\ -\epsilon /2 ~[{\mathbf J}]_{i,j}, & \text{if } i=j \text{ and } k_1 \neq k_2  \\ 0, & \text{otherwise} \end{matrix} \right\} \text{.}
    \label{MultiJ}
\end{multline}
D-Wave provides a solver which is used to pre-compute an embedding, given the initial entries of $\hat{\mathbf J}$ to be used. Using this, we define the qubits, couplers, and chains on the processor to solve a problem of a given size. Each unique embedding for different $K$ and $N$ is stored in memory for the time when a problem of size $K$ and $N$ needs to be solved.

\begin{table}[b]
\caption{Computational Cost of PCA Algorithms}
\label{table}
\setlength{\tabcolsep}{3pt}
\begin{tabular}{|p{90pt}|p{160pt}|}
\hline
Type& 
Computational Cost \\
\hline
SVD \cite{golub1971singular}& 
$\mathcal O(\min\{N,D\}ND)$\\
L1-BF \cite{markopoulos2017efficient}& 
$\mathcal O(ND\min\{N,D\}+N^2K^2(K^2 + d))$\\% + N^2d
QAPCA (ours)& 
$\mathcal O(N^2 D+K^2(N-\hat{\kappa})^2)$\\
QAPCA-R (ours)& 
$\mathcal O(K N^2 D+K(N-\hat{\kappa})^2)$\\
\hline
\multicolumn{2}{p{225pt}}{}\\%
\end{tabular}
\label{tabComputationalCost}
\end{table}

\subsection{Complexity Analysis}
The time to perform QAPCA is driven by many of the classical computing elements found in problem pre-processing and post-processing. Initialization of the problem will involve calculation of matrix $\mathbf J=-\mathbf X^\top \mathbf X$ which costs $\mathcal O(N^2 D)$. Calculation of modified matrix $\hat{\mathbf J}$ is done in a few steps. The first step is the operation of halving the scale of the diagonal elements has cost $\mathcal O(N)$. Scaling every element in $J$ to ensure it can be represented on the annealer will have cost $\mathcal O(N^2)$. The banding operation which involves finding $\kappa$ then loading only the associated elements into $\hat{\mathbf J}$ will cost $\mathcal O(\hat{\kappa} + \frac{(N-\hat{\kappa})(N-\hat{\kappa}+1)}{2})= \mathcal O(\hat{\kappa}+(N-\hat{\kappa})^2)$. Assigning values to each of the couplers on the annealer in the embedding process costs $\mathcal O(\frac{K(K+1)}{2}\frac{(N-\hat{\kappa})(N-\hat{\kappa}+1)}{2})=\mathcal O(K^2(N-\hat{\kappa})^2)$. 

The annealing process itself will only cost the complexity of outputting the binary matrix $\mathcal O(aKN)$, here $a$ is the amount of times the problem is annealed.
Post-processing involves calculation of the principal components from the binary output costs $\mathcal O(KND+KD)=\mathcal O(KND)$. Then, the $\Phi (\cdot)$ operation from \eqref{Procrustes} costs $\mathcal O(KD^2+KD^2)=\mathcal O(KD^2)$.

The complexity of the QAPCA process is dominated by the terms
\begin{multline} 
\mathcal O(N^2 D)+\mathcal O(\hat{\kappa}+(N-\hat{\kappa})^2)+\mathcal O(K^2(N-\hat{\kappa})^2)+\mathcal O(aKN)\\+\mathcal O(KND)+\mathcal O(KD^2).
\end{multline} 
The method's complexity is quadratic in $N$ and $K$ and $D$.

 When computing QAPCA-R the complexity is very similar except that we must account for the performing the operation of nullspace projection $\mathcal O(KND^3)$, and the fact the entire process must be repeated $K$ times. The complexity of performing the QAPCA process with nullspace projection is dominated by the terms
\begin{multline} 
\mathcal O(KN^2 D)+\mathcal O(K^2ND^3)+\mathcal O(K\hat{\kappa}+K\hat{\kappa}(N-\hat{\kappa})^2)+\mathcal O(K(N-\hat{\kappa})^2)\\+\mathcal O(aK^2N)+\mathcal O(KND).
\end{multline} 
The cost of performing the nullspace projection $K$ times is cubic in $D$ and quadratic in $N$ and $K$. Table \ref{tabComputationalCost} shows that the computational cost of QAPCA is competitive with that of the Bit Flipping algorithm (L1-BF) \cite{markopoulos2017efficient} when there are multiple components to calculate.

QAPCA can show a linear speedup of $K^2$ when compared to L1-BF if $d$ is the true rank of the dataset.

\section{Numerical Studies}
We now explore the performance of the QAPCA algorithm in studies where each dataset is known to have a relatively low number of features. We compare the results to the standard (L2-norm) PCA, computed by classical SVD and L1-PCA, computed by means of the L1-BF algorithm \cite{markopoulos2017efficient}. The annealing process is repeated 10 times for every output of QAPCA and only 5 times for each unique component of QAPCA-R. The $\hat{\mathbf b'}$ which minimizes \eqref{MultNucNorm6} is selected. The classical portion of the following simulations was run on an AMD Ryzen 5950X at 4000MHz using 8GB of RAM.

%We will use the value of the L2 norm to determine the classification of a test sample. 

%If $\|X^{\top}R_{Malignant}\|>\|X^{\top}R_{Benign}\|$ Classify as Malignant, otherwise classify as Benign.

\begin{figure}[b!]
	\centering
	\includegraphics[width=3.3in]{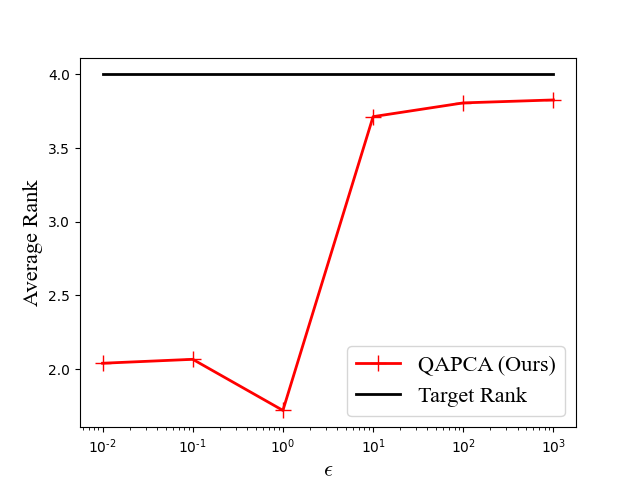}
	\caption{Average Rank vs. $\epsilon$, $N=20$, $K=4$ of algorithms on Gaussian example problem before \eqref{Procrustes} is applied.}
	\label{MULTIAR126}
\end{figure}

\begin{figure}[t!]
	\centering
	\includegraphics[width=3.3in]{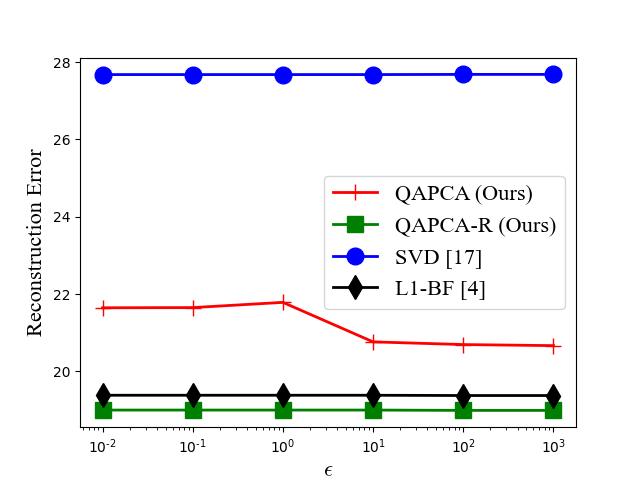}
	\caption{Reconstruction Error vs. $\epsilon$, $N=20$, $K=4$ of algorithms on Gaussian example problem. Red indicates QAPCA, Green indicates QAPCA-R, blue indicates SVD, and black indicates L1-BF.}
	\label{MULTIRE126}
\end{figure}

\begin{figure}[t!]
	\centering
	\includegraphics[width=3.3in]{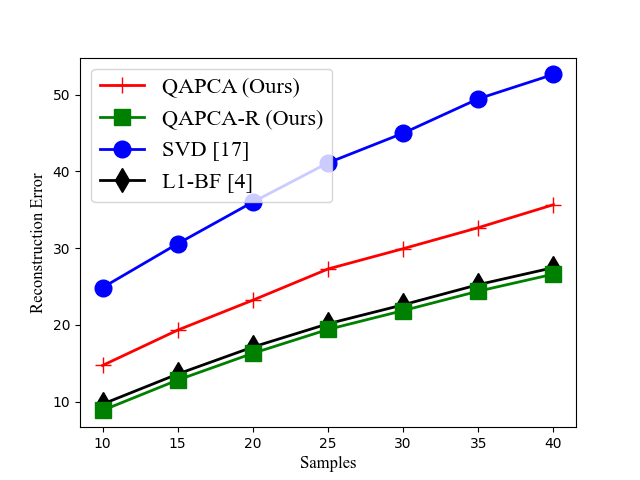}
	\caption{Reconstruction Error vs. $N$, $K=4$, $\epsilon=100$ of algorithms on WBCD.  Red indicates QAPCA, Green indicates QAPCA-R, blue indicates SVD, and black indicates  L1-BF.}
	\label{MULTIRE125}
\end{figure}

\subsection{Gaussian Toy Problem}
A scenario is conceived where samples are drawn from the superposition of three multivariate Gaussian distributions each with $D=50$. During each run of our evaluation we create new Gaussian distributions with mean and covariance entries uniformly drawn from $[-1,1]$, the third Gaussian has a biased covariance with $-9$ added to every entry. The samples are superimposed than set to have zero mean and scaled to have unit standard deviation. Results for the figures were averaged over 100 different realizations. 

The number of unique components in QAPCA before \eqref{Procrustes} is performed is shown in Fig. \ref{MULTIAR126}. As $\epsilon$ increases, QAPCA is more likely to find components orthogonal to the other components. New components are more likely to occur once we have an $\epsilon > 1$.

The reconstruction error in Fig. \ref{MULTIRE126} shows that L1-PCA performs better than L2-PCA when there is a large contributor to overall variance. QAPCA performs better than L2-PCA but not as well as L1-BF. QAPCA reconstruction error is notably lower after $\epsilon > 1$ this is due to the increased number of unique components being generated. When the QAPCA-R is used, the performance is comparable to that of L1-BF.

\begin{figure}[t!]
	\centering
	\includegraphics[width=3.3in]{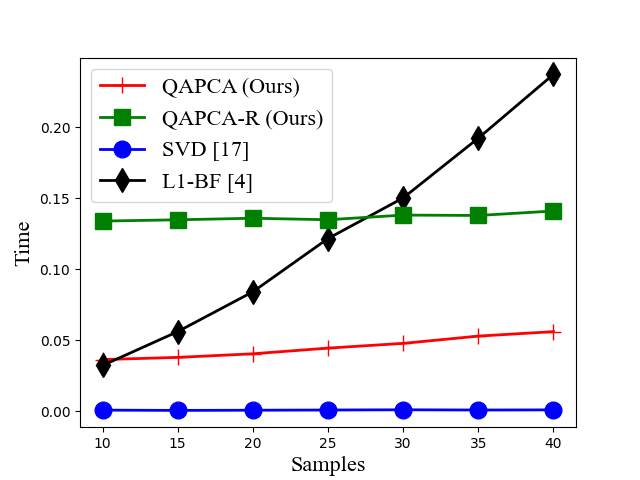}
	\caption{Time vs. $N$, $K=4$, $\epsilon=100$ of algorithms on WBCD. Red indicates QAPCA, Green indicates QAPCA-R, blue indicates L2-PCA using SVD, and black indicates L1-PCA using L1-BF.}
	\label{MULTITime}
\end{figure}

\subsection{Wisconsin Breast Cancer Diagnostic}

In this experiment, we mislabel data from the Wisconsin Breast Cancer Diagnostic (WBCD) dataset \cite{wolberg1992breast}. After centering the all of the data on the median and scaling all of the data between the $1_{\text{st}}$ and $3_{\text{rd}}$ quartile to be within $[-1,1]$, we intentionally mislabel 20\% of the data that the PCA is being trained upon. To ensure unique components, we have set $\epsilon = 100$. Results for the figures were averaged over 100 different realizations. 
%There appears to be a performance increase after we reach $\epsilon = 1$, possibly due to some data in $\mathbf X$ being outside of $[-1,1]$.

The results in Fig. \ref{MULTIRE125} again show L1-PCA with better reconstruction than L2-PCA in the presence of mislabeled data. QAPCA shows improvement over L2-PCA but never reaches the performance of the L1-BF whereas QAPCA-R performs just as well as the L1-BF. 

Fig. \ref{MULTITime}, shows the execution time for the various algorithms in seconds. Unexplained transmission and receive delays have been omitted from the timing results of the quantum annealer. QAPCA shows the time the annealer took on the problem plus the calculation and translation time to convert $\mathbf J$ to a list of embedded couplers plus the time it takes to calculate the PCA given the returned optimal bits. In Fig. \ref{MULTITime} we see L2-PCA and QAPCA can operate much faster than L1-BF. QAPCA can increase the speed of L1-PCA by orders of magnitude while still remaining robust during training. We observe the linear time penalty as QAPCA must perform coupler assignment for larger Ising problems. The coupler assignment penalty increases as $N$ increases. We can also start to see a time benefit of using QAPCA-R in comparison to L1-BF.

\subsection{Tennessee Eastman Problem}
We now explore the performance of QAPCA-R in a fault detection study where the data has a relatively low number of features. We introduce outliers to the Tennessee Eastman Problem (TEP) \cite{downs1993plant} where the data has been sourced from \cite{DVN/6C3JR1_2017}. To create outliers, Gaussian noise with a standard deviation of 100 
was randomly added to 20\% of the data. For increased accuracy, the L1-BF has been run over 32 different initializations run in parallel. Results for the figures were averaged over 50 different realizations.

A common metric for fault detection is the Squared Predicted Error (SPE) \cite{macgregor1995statistical}. otherwise known as the $Q$ statistic. The metric for a test sample $\mathbf x_i \in \mathbb R^{D}$ can be calculated as
\begin{align}
 \text{SPE} = \|\mathbf x_i-\mathbf R \mathbf R^\top \mathbf x_i\|_2^2,
\end{align}
where $\|\cdot\|_2$ is the Euclidean norm of a vector.

The number of true negatives and false positives are calculated from the combination of a faultless test set (Fault 0) and up to sample 160 in the faulty datasets for each associated faulty test set from a total of $N_\text{faultless} = 4160$ samples. If we have $N_\text{false+}$ as the number of false positives, then the false positive rate (FPR) is calculated as
\begin{align}
\text{FPR} = \frac{N_\text{false+}}{N_\text{faultless}}.
\end{align}
The number of true positives and false negatives are counted after sample 160 for each associated faulty test set from a total of $N_\text{faulty} = 16000$ samples. If we have $N_\text{true+}$ as the number of true positives, then the true positive rate (TPR) or recall is calculated as
\begin{align}
\text{TPR} = \frac{N_\text{true+}}{N_\text{faulty}}.
\end{align}
% We will also 
We calculate the Precision, which is defined to be 1 if $N_\text{false+}+N_\text{true+} = 0$ but otherwise is
\begin{align}
\text{Precision} = \frac{N_\text{true+}}{N_\text{false+}+N_\text{true+}}.
\end{align}

The ROC curves are created by plotting FPR versus TPR at different detection thresholds and the PRC curves are created by plotting TPR versus Precision. 
To form smooth curves, the detection limits have been evaluated at  points between [0, 10e-5] with steps of 10e-6, between [10e-5, 10e-4] with steps of 10e-5, between [10e-4, 0.5] with steps of 10e-4, between [0.5, 5] with steps of 10e-3, between [5, 300] with steps of 1, between [300, 1000] with steps of 100, and between [1000, 10e3] with steps of 1000. The Area Under the Receiver Operating Characteristic curve (AUROC) and the Area Under the Precision-Recall Curve (AUPRC) are metrics less biased than metrics with set detection limits.

\begin{figure}[t!]
	\centering
	\hspace*{-1in}
	\includegraphics[width=6.5in]{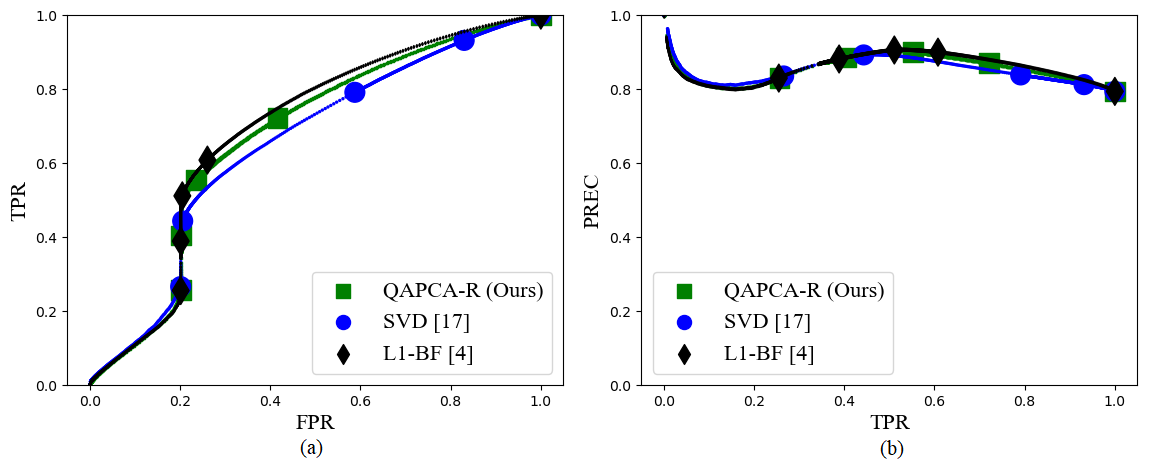}
	\caption{Fault detection for $N=150$, $K=20$ of on TEP. Green indicates QAPCA-R, blue indicates SVD, and black indicates L1-BF.(a) ROC. (b) PRC.}
	\label{TEP_ROC_PRC}
\end{figure}

\begin{figure}[t!]
	\centering
	\includegraphics[width=3.3in]{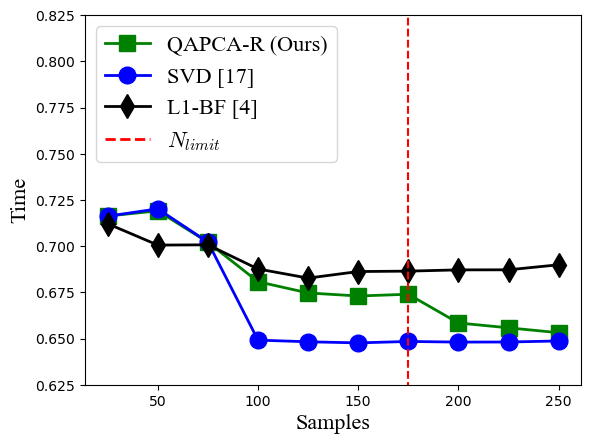}
	\caption{Performance of AUROC vs. $N$, $K=20$ of algorithms on TEP. Green indicates QAPCA-R, blue indicates SVD, and black indicates L1-BF.}
	\label{AUROC_N_234}
\end{figure} 
 
\begin{figure}[t!]
	\centering
	\hspace*{-1in}
	\includegraphics[width=6.5in]{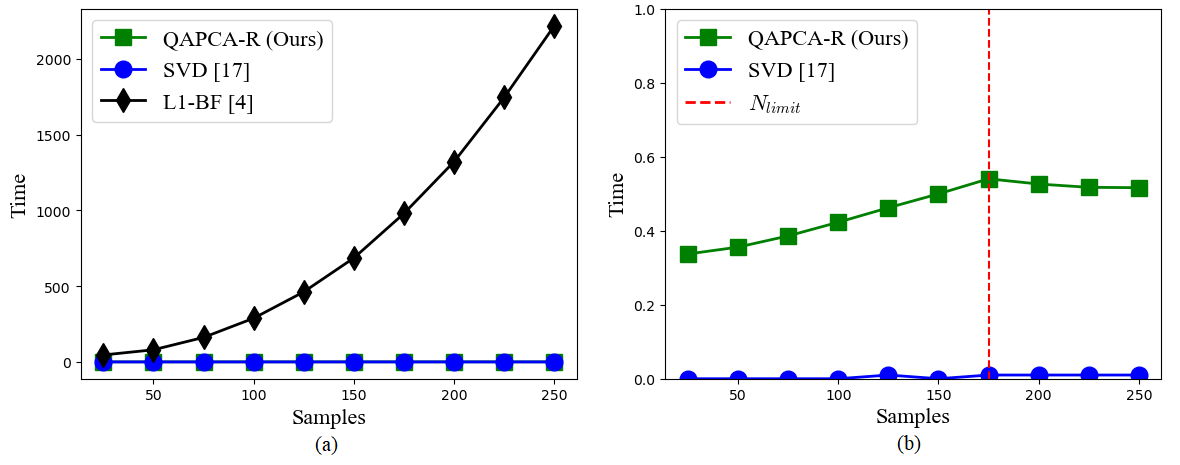}
	\caption{Time vs. $N$, $K=20$ of algorithms on TEP. Green indicates QAPCA-R, blue indicates SVD, and black indicates L1-BF. (a) Time L1-BF. (b) Time $< 1$s.}
	\label{TEP_time_234}
\end{figure}

The fault detection results in Figure \ref{TEP_ROC_PRC}a show that L1-BF performs better than SVD once the false positive rate is higher than 20\%.

This is the point where the detection limits are low enough to account for the percentage of outliers being added, or the region where SPE detection is actually useful. When this occurs it causes the TPR to jump to around 40\% for SVD. In Figures \ref{TEP_ROC_PRC}a and \ref{TEP_ROC_PRC}b, once the TPR reaches 40\% L1-BF begins to outperform SVD. QAPCA-R is not able to perform as well as L1-BF but it still outperforms SVD.

The results in Figure \ref{AUROC_N_234} show that SVD outperforms L1-BF when the number of samples is low, as the number of outliers is low as well. However, SVD cannot be calculated for lower than $K$ samples, so using SVD is only valid in a small range. QAPCA-R matches SVD's performance in this range. The L2 metric gives a larger error response to faults and outliers in general. Once the number of outliers is great enough to severely corrupt the data, L1-BF begins to outperform SVD. This is because L1-BF is robust to outliers and a better set of principal components is generated. QAPCA-R remains a strong choice until the banding embedding procedure is required to fit the samples on the annealer. QAPCA-R still outperforms SVD in this range but eventually the QAPCA-R performance degrades to that of SVD.

Figure \ref{TEP_time_234}, shows the execution time for the various algorithms in seconds. Unexplained transmission and receive delays have been omitted from the timing results of the quantum annealer. QAPCA-R is the time the annealer took on the problem plus the calculation and translation time to convert $\mathbf J$ to a list of embedded couplers plus the time it takes to calculate the PCA given the returned optimal bits. In Figure \ref{TEP_time_234}a we see SVD and QAPCA-R can operate much faster than L1-BF.  QAPCA-R can increase the speed of  L1-BF by orders of magnitude while still remaining robust during training. QAPCA-R is rerun for every new component, whereas L1-BF and SVD only need to perform this operation once. In Figure \ref{TEP_time_234}b we observe the compounded time penalty as QAPCA-R must repeat the coupler assignment and the annealing procedure for each new component. The coupler assignment penalty increases as $N$ increases, until we begin to implement banding at $N=175$ wherein only $C_\text{limit}$ couplers are assigned.

\section{Conclusions}
\label{sec:Future Work}
In this paper we introduced QAPCA and demonstrated that quantum annealing is a viable approach for accelerating L1-PCA. 
Our QAPCA algorithm is able to outperform L2-PCA in component quality in the presence of mislabeled data. This motivates interest in quantum hardware for the field of classification and dimensionality reduction facilitating various potential applications. In the future this method can be optimized for various levels accuracy while benefiting from increased speed.

% References
\bibliography{refs} % bibliography data in report.bib
\bibliographystyle{spiebib} % makes bibtex use spiebib.bst

\end{document}